\documentstyle[aps,epsf,amssymb,prl]{revtex}

\begin{document}

\title{Pseudo-contact angle due to superfluid vortices in $^{4}$He }
\author{R. Luusalo, A. Husmann \thanks{
Present address: Clarendon Laboratory, 
Oxford University, Oxford OX1 3PU, England}, J. Kopu, and P.J. Hakonen}
\address{Low Temperature Laboratory, Helsinki University of Technology, Espoo,
P.O.\\
Box 2200, FIN-02015 HUT, Finland}
\date{\today }
\maketitle

\begin{abstract}
We have investigated spreading of superfluid $^{4}$He on top of polished 
MgF$_2$
and evaporated SiO$_{2}$ substrates. Our results show strongly varying contact
angles of 0 - 15 mrad on the evaporated layers. According to our
theoretical calculations, these contact angles can be explained by a
spatially varying distribution of vortex lines, the unpinning velocity of
which is inversely proportional to the liquid depth.
\end{abstract}


\bigskip 



Wetting phenomena in superfluid $^{4}$He have been under intense
investigation during the past few years. After the theoretical prediction by
Cheng et al. \cite{cheng:cesium}, it was found out by Nacher and 
Dupont-Roc \cite
{nacher:wetting} that, indeed, the alkaline metal Cs is not wetted by
superfluid $^{4}$He. Recently, Klier et al. \cite{klier:cesium} measured $%
\theta =48^{\text{o}}$ for the contact angle on top of an evaporated Cs film
using capillary rise methods. Subsequent optical experiments \cite
{ross:spreading,rolley:roughness} gave slightly smaller values $\theta
=25-32^{\text{o}},$ and they also revealed strong hysteresis between
advancing and receding liquid fronts. This hysteresis cannot be fully 
explained in terms of external disorder, although roughness of the advancing
contact line has been found to agree with classical scaling 
laws \cite{rolley:roughness}.

Similarly, hysteretic variation of small contact angles was observed in
optical experiments where spreading on top of a commercial antireflection
coating was investigated \cite{alles:wetting}. In this Letter we report a
detailed investigation on this kind of contact lines and present results,
both theoretical and experimental, which show that a high density of pinned
superfluid vortices leads to apparent contact angles of a few degrees in the
surface profile. In our experiments at 1 K, such contact angles were visible
on top of evaporated SiO$_{2}$ layers but not on a smooth, bulk MgF$_{2}$
substrate. The difference follows from the larger amount of pinning sites
for superfluid vortices on the evaporated films than on a polished surface. 
Our results suggest strongly that intrinsic disorder, {\it i.e.} pinned 
vortices cannot be neglected when considering the contact line dynamics 
of $^{4}$He-II, {\it e.g.} on evaporated Cs films.

Vortices are always present in thin $^{4}$He superfluid films; even in bulk
liquid it is difficult to reach a vortex-free state \cite
{awschalom:pinning,hakonen:pinning}. This is attributed to the fact that
vortex core radius $a_{0}$\ is about 1 \AA\ \cite{donnelly:book} in $^{4}$He-II
and, hence, pinning sites of atomic size $b$ can trap vortices strongly. Due
to the atomic size, a large number of pinning sites can be
found on the surface, easily up to densities on the order of 
$10^{16}$ m$^{-2}$.

Unpinning of vortices in a thickening film is caused by the interaction
between neighboring vortices. A vortex will adjust itself to the flow field
caused by an adjacent vortex line by bending \cite{schwarz:pinning}. Once
its radius of curvature is smaller than the liquid layer thickness $h$, the
vortex becomes unpinned. Using a hemispherical pinning site of radius \ $b$,
Schwarz calculated the unpinning velocity $v_{0}$ numerically \cite
{schwarz:pinning}. His results can be summarized by 
\begin{equation}
v_{0}=(\kappa _{4}/4\pi h)\ln (b/a_{0})\;\;,  \label{vd}
\end{equation}
where $\kappa _{4}=2\pi \hbar /m_{4}$. Since the pinning potential is deep, $%
\sim 100$ K for a typical atomic site of $b=1$ nm, thermally activated
unpinning at 1 K is effective only in the vicinity of $v_{0}$ \cite
{schwarz:thermaldep} and the geometric law of Eq. (\ref{vd}) is a good
approximation in our region of interest. If we assume that unpinning is
caused by a single nearby vortex inducing a velocity $v=(\kappa _{4}/2\pi r),
$ then we obtain 2$h$ for the minimum intervortex distance within
logarithmic accuracy. Hence, the maximum areal density $n_{0}$ of vortices
in a spatially varying film can be estimated by the relation \cite{estimate} 
\begin{equation}
n_{0}(\overrightarrow{R})=\frac{1}{4h^{2}(\overrightarrow{R})}\;\;,
\label{density}
\end{equation}
where $\overrightarrow{R}$ denotes the position in the plane of the
substrate. 

The free energy of a superfluid film with a vortex density can be written as

\begin{equation}
E_{s}=\int_{V} n_{\rho} U(\overrightarrow{r})dV+\int_{S}\sigma dS\;\;,
\label{energy}
\end{equation}
where $n_{\rho}$ is the particle density and $\sigma $ is the surface
energy per unit area. The potential $U(\overrightarrow{r})$ denotes the
energy per atom of mass $m_{4}$: 
\begin{equation}
U(\overrightarrow{r})=m_{4}gz-\frac{\gamma(z)}{z^{3}}
+\frac{1}{4\pi}\frac{\rho _{s}}{n_{\rho}}\kappa _{4}^{2}n_{v}
(\overrightarrow{R})\ln (l/a_{0})\;\;,
\label{potential}
\end{equation}
where $\gamma (z)$ is a function governing the strength of the van der Waals
interaction (vdW), which decreases towards higher $z$ due to the retardation
effects that are known to
be important for distances larger than about 5 nm \cite{isr:surf}. 
The last term denotes the kinetic
energy due to vortices with superfluid density $\rho _{s}$. The distance $l$
is the upper cut-off for the vortex flow field that can also be expressed as 
$\simeq 1/2\sqrt{n_{v}(\overrightarrow{R})}$. We set the logarithmic term equal
to one in our calculation, which means that we underestimate the influence
of vortices a bit. We also neglect the retardation effects. 
The vdW term in Eq. (\ref{energy}) is approximated by the interaction
energy between two flat surfaces, $A/12\pi h^{2}$ per unit area, where $A$
is the conventional Hamaker constant.

The minimization of Eq. (\ref{energy}) depends crucially on the behavior of
vortex density with increasing film thickness. When $n_{v}(\overrightarrow{R}%
)$ is below the critical unpinning limit given by Eq. (\ref{density}), the
kinetic energy of vortices $E_{s}^{vort}(h)$ $\propto h$. On the other hand,
when vortices start to become unpinned and their density follows the maximum
value of $1/4h^{2}$, then $E_{s}^{vort}(h)$ $\propto 1/h$ and, in fact, the
gradual elimination of vortices by unpinning favors thickening of the film.
This behavior leads to two stable minima in $E_{s}(h)$ as is illustrated in
Fig. 1 without surface tension.

The lower local minimum at film thickness $h_{1}$ is given by the
competition between vdW forces and the elastic tension of vortices, which
leads to the relation

\begin{equation}
h_{1}=\left[ \frac{2A}{3\rho _{s}\kappa _{4}^{2}\ln (l/a_{0})n_{v}}
\right] ^{1/3}\;\;\;.  \label{min1}
\end{equation}
By setting $n_{v}$ to its upper limit $n_{0} = 1/4h_{1}^{2}$, Eq. (\ref{min1}) 
can be used to solve for
the minimum thickness $h_{1}=h_{t}$ below which no vortex-thinned 
film can exist. When 
$h<h_{t}$, the van der Waals force wins over the vortex tension 
caused by any vortex density below the limit of Eq. (\ref{density}).
In other words, the vdW attraction will always limit the vortex
density below the value of $n_{0}^{crit}=1/4h_{t}^{2}$. Using $\ln
(l/a_{0})=1$, $\rho _{s}=140$ kg/m$^{3}$, and $A=-5.9\cdot 10^{-21}$J
measured for CaF$_{2}$ \cite{sabisky:films}, we obtain $h_{t}=12$ nm and $%
n_{0}^{crit}=1.9\cdot 10^{15}$ m$^{-2}$.

At a larger thickness, another minimum is obtained when the decrease of energy
due to elimination of vortices is balanced by an increase in the
gravitational energy, which yields the value  
\begin{equation}
h_{2}=\left[ \frac{\kappa _{4}^{2}\rho _{s}\ln (l/a_{0})}{16\pi g \rho}\right]
^{1/3}\;\;\;,  \label{min2}
\end{equation}
when $n_{v0} > 3\cdot 10^{10}$ m$^{-2}$.
Here we assume that there is a fluid
reservoir at the level of the substrate. Note that the above value is much
larger than the thickness of a regular vdW film because the vortex energy
dies away as $1/h.$ \ A transition between thin (Eq. \ref{min1}) and thick
(Eq. \ref{min2}) films leads to an abrupt kink in the slope, reminiscent of
a contact angle in the surface profile. Since the substrate is completely
wet, we call this edge a pseudo-contact angle in order to make a clear
distinction with a true three phase contact line.

In our numerical calculations the one-dimensional surface profile 
$h(x)$ was taken
to be defined in $N$ equidistant points $x_{1},...,x_{N}$ along the substrate.
Then the minimum of the discretized free energy $E(h_{1},...,h_{N})$, where
$h_{i}=h(x_{i}$), was determined numerically with standard methods under the
requirement that the total volume of the fluid remains constant. 
For the surface tension we used the zero-temperature
value of $\sigma $ = 375 $\mu $J/m$^{2}$ \cite{roche:tension}.

Fig. 2 illustrates the behavior of different energy terms when a cross-over
from a thin to a thick film takes place. On the thin film side the
vortex density is taken to be $n_{v0}=10^{13}$ m$^{-2}$, which
corresponds to a unpinning 
thickness of 160 nm. For films thicker than that,
$n_{v}(\overrightarrow{R})$ follows its maximum value  $n_{0}$ given by 
Eq. (\ref{density}).
First, when the film thickness increases from 70 nm,
the flow energy grows quickly ($\propto h$), followed by a strong increase
in the surface energy and by a rapid decrease in the vdW energy. The kinetic
energy reaches its maximum at the thickness when vortices start to become
unpinned. Above the unpinning threshold, the surface energy gradually
relaxes to zero, while the vortex energy is reduced to a value balanced by
the gravitational potential of the fluid. The inset displays the corresponding
surface profile. An angle of 31 mrad is visible as the film breaks
off from the substrate. 
For the pseudo-contact angle, we take
the maximum slope of the surface with respect to the substrate.

The calculated pseudo-contact angle $\theta_{p}$ as a function of the vortex
density $n_{v0}$ is displayed in Fig. 3. 
Results at small $n_{v0}$ calculated on
a substrate inclined by 3.5 mrad from the horizon show that
the pseudo-contact angle goes smoothly to the value of the substrate
inclination, {\it i.e.}, there is no observable contact angle at small vortex
densities  $n_{v0} \sim 10^{10}$ m$^{-2}$. The calculations on horizontal
substrate, on the other hand, indicate a discontinuous vanishing
of $\theta_{p}$ at $n_{v0} = 3\cdot 10^{10}$ m$^{-2}$.

Our experiments were performed on a dilution refrigerator equipped with a
two-beam interferometer; for details we refer to Ref. \cite{ruutu:long}. The
investigated substrates were inclined by $\sim $5 mrad from the
horizon. We made experiments on three
different substrates: 1) Bulk MgF$_{2}$, 2) AR-coating of $Hebbar$-type, and
3) AR-coating of $V$-type \cite{melles-griot}.  In our
measurements at 1 K and below, bulk MgF$_{2}$ was always covered by a smooth 
$^{4}$He-II film whereas substrates 2 and 3 were not; we limit our
observations to low temperatures because, in this range, the
regulation of liquid level
worked well and stable fluid fronts could be obtained.

Wetting behavior on top of the polished MgF$_{2}$ substrate is illustrated in
the upper part of Fig. 4. It displays a smooth surface profile, deduced from
the interferogram shown in the inset. The shape of the surface can be
accounted for by the regular van der Waals film formula as can be seen 
from the fit to
the data. This kind of a film is well in accordance with the measurements by
Sabisky and Anderson using CaF$_{2}$ and SrF$_{2}$ substrates \cite
{sabisky:films}. The lower part of Fig. 4 illustrates the behavior of $^{4}$%
He-II on top of evaporated $Hebbar$-coating. A distinct
pseudo-contact angle of 6 mrad is observed in the profile.
Similar surface shapes were also seen on top of the $V$-coating. 

Strong hysteresis between advancing and receding front was observed on the $V
$-coating as well as on the $Hebbar$-coating. The contact angle was found to
be zero for receding liquid whereas the advancing front displayed
pseudo-contact angles in the range of 0 - 15 mrad. Moreover, when shaking the
cryostat, the liquid left behind tracks which were preferred by the next
advancing front. All these features are nicely explained by a variation in
the density of trapped vorticity in thin superfluid films.

According to Fig. 3, the experimentally measured pseudo-contact angles of 1
- 15 mrad correspond to $n_{v0}= 10^{10} - 5\cdot 10^{11}$ m$^{-2}$. These
densities are clearly larger than the values measured by Ellis and Li \cite
{ellis:films} who got values on the order of 10$^{9}$ m$^{-2}$ in their
third sound measurements. However, all our densities are much smaller
than the physical maximum value $1.9\cdot $10$^{15}$ m$^{-2}$ limited
by the vdW attraction (see Eq. (\ref{min1})).

It has been suggested by Williams and Wyatt \cite{williams:charge} that
surface charge might play a role in helium spreading on top of an
AR-coating. Because of the polarizability of helium atoms, the energy
density in the areas exposed to an electric field $E_{0}$ decreases by $%
\Delta E_{\rm liq}=\frac{1}{2}\chi\epsilon _{0}E_{0}^{2}$ where $\chi $
is the dielectric susceptibility. If strong enough electric fields with sharp
spatial gradients were formed, the surface profile could display distinct
borders with pseudo-contact angles. However, the irregular behavior observed
in our experiments is against the model of Williams and Wyatt: if frozen
charge is important then a rather reproducible pattern should be seen when
the liquid front is taken back and forth over the substrate.

Our substrates were checked for uniformity using an atomic force
microscope (AFM). The roughness was found to be 0.4, 1.0, and 1.0 nm$_{rms}$ 
for
samples 1 - 3, respectively. In addition, morphology of the roughness
was found to be different: for sample 1, the roughness was mostly
caused by long, continuous scratches, while samples 2 and 3 showed
irregular, granular-like patterns. Hence uniform, ``granular''
roughness on mesoscopic scale
appears to favor vortex pinning as expected. Moreover, we used the AFM 
to deposit surface charge and to follow directly its decay. We found
that the lifetime of charge on our substrates was on the order of ten
minutes, which was far too short to leave any appreciable amount of charge
behind after the mounting and cool-down procedures lasting for several hours.

Recently, Herminghaus \cite{herminghaus:droplet} considered the effect
of Bernoulli pressure on contact
lines. His calculation basically applies to the
spreading situation where vortex nucleation limits the superflow and
thereby governs the spreading dynamics. He obtained contact angles on the
order of 5 degrees. However, in order to get stable liquid fronts with a
finite contact angle in this way, a continuous heat flux should be present
in our experiments, which is not the case. Moreover, the pattern of the heat
flux should be a rather complicated, time-dependent function of the position
on the substrate.

On the basis of the present experiments and calculations, we believe that
the elastic energy of pinned vortices influences spreading of helium on top
Cs \cite{klier:cesium,ross:spreading,rolley:roughness}, causing extra 
hysteresis in the indigenous contact angle and pinning of the contact line. 
In the case of Cs, even an
order of magnitude larger density of vortices is possible, since the
evaporation of metal is done at low temperatures in order to avoid
oxidation. Plenty of pinning sites were observed in recent flow experiments 
\cite{hakonen:pinning} where solid air or solid hydrogen had been evaporated
at low temperatures.

To conclude, quantized vortices play a role in the spreading of superfluid
when there are plenty of pinning sites for vortices. The elastic tension of
vortex lines leads to thin film sections that look like non-wetted regions
even though they are covered by $^{4}$He-II. Our calculations on
pseudo-contact lines at the borders of these thin film regions are in good
agreement with experimental results obtained on evaporated antireflection
coatings. Owing to unpinning of vortices in thick fluid layers, strong
hysteresis with respect to advancing and receding liquid front is produced.
This finding corroborates with our results, as well as recent experiments on
Cs films. Hence, the dynamics and pinning of a superfluid contact line can
be strongly influenced by vortices on atomically rough substrates.

We want to thank S. Balibar, M. Paalanen, E. Rolley, and E. Sonin for useful
discussions and correspondence. This work was supported by the Academy of
Finland and by the Human Capital and Mobility Program ULTI of the European
Community.

\begin{figure}[tbp]

\caption{Energy per unit area {\it vs.} the thickness $h$ of the superfluid
layer. Total energy neglecting surface tension is drawn by the solid line.
The substrate level is taken as zero for the
gravitational energy, the vdW energy is zero at $z=\infty$, and the vortex 
contribution is calculated for the
density $n_{v0}=10^{13}$\ m$^{-2}$. For other parameter values, see text. The
two minima $h_{1}$ and $h_{2}$ given by Eqs. (\ref{min1}) and (\ref{min2}),
respectively, exist only when $3\cdot 10^{10}$\ m$^{-2}<n_{v0}<1.9\cdot
10^{15}$\ m$^{-2}$.}
\label{f1}
\end{figure}

\begin{figure}[tbp]

\caption{Energy densities and the surface profile
(inset) calculated at $n_{v0}=10^{13}$ m$^{-2}$ across
the thin-to-thick film transition region. The solid curve is 
surface tension, dash-dotted is vortex energy and the dashed
curve is the van der Waals interaction energy. The
maximum slope yields the pseudo-contact angle 31 mrad.}
\label{f2}
\end{figure}

\begin{figure}[tbp]

\caption{Pseudo-contact angle $\theta_{p}$ {\it vs.} vortex density $%
n_{v0}$ calculated for substrate inclination of 0 ($\circ$) and 
3.5 mrad ($\bullet$).}
\label{f3}
\end{figure}

\begin{figure}[tbp]

\caption{Interferograms and the corresponding surface profiles measured on
bulk MgF$_{2}$ substrate (upper frame) and on {\it Hebbar}-coating (lower
frame); the dashed lines denote the substrates inclined from the horizon. The
profiles have been calculated along the white lines marked in the
interferograms. The surface on top of MgF$_{2}$ fits the standard vdW
film formula illustrated by the solid curve. On the {\it Hebbar}-coating a
pseudo-contact angle $\protect\theta _{p}=6$ mrad is visible.}
\label{f4}
\end{figure}










\begin{references}
\bibitem{estimate}  See also Ref. \cite{awschalom:pinning} which gives a
larger estimate than ours.
\end{references}
\end{document}